# Exact solution of the 1D Dirac equation for the inverse-square-root potential $1/\sqrt{x}$

**A.M. Ishkhanyan[1,2]**

[1]Russian-Armenian University, Yerevan, 0051 Armenia
[2]Institute for Physical Research, NAS of Armenia, Ashtarak, 0203 Armenia

We present the exact solution of the 1D Dirac equation for the inverse-square-root potential $1/\sqrt{x}$ for several configurations of vector, pseudo-scalar and scalar fields. Each fundamental solution of the problem can be written as an irreducible linear combination of two Hermite functions of a scaled and shifted argument. We derive the exact equations for bound-state energy eigenvalues and construct accurate approximations for the energy spectrum.

**1. Introduction**

Exact solutions of the Dirac equation, which is a relativistic wave equation of fundamental importance in physics [1], are rare. In the one-dimensional case, apart from the piece-wise constant potentials and their generalizations involving the Dirac $\delta$-functions, one may mention the Coulomb, linear, and exponential potentials [2]

$$V = V_0 + \frac{V_1}{x}, \quad V_0 + V_1 x, \quad V_0 + V_1 e^{x/\sigma}, \tag{1}$$

for which the Dirac equation can be solved in terms of confluent hypergeometric functions, and the potential

$$V = V_0 + V_1 \tanh\left(\frac{x}{\sigma}\right) + V_2 \coth\left(\frac{x}{\sigma}\right), \tag{2}$$

for which the Dirac equation can be solved in terms of ordinary hypergeometric functions. Many potentials reported so far, e.g., the Dirac-oscillator [3,4], Woods-Saxon [5], and Hulthén [6] potentials are particular cases of these four potentials that can be derived by (generally complex) specifications of the involved parameters. Note that the first three potentials given by equation (1) are particular truncated cases of the classical Kratzer [7], harmonic oscillator [8], and Morse [9] potentials for the Schrödinger equation, while the fourth potential given by equation (2) in general (if $V_1 V_2 \neq 0$ and $V_1 \neq V_2$) does not belong to a known Schrödinger potential. However, the solution for this potential can be constructed using the solution of the Schrödinger equation for the Pöschl-Teller potential [10] or that for the Eckart potential [11] (note that the solution for a truncated version of potential (2) with $V_2 = 0$ can be written using the solution for the Eckart potential).

In the present paper, we introduce a new exactly solvable potential – the inverse-square-root potential

$$V = V_0 + \frac{V_1}{\sqrt{x}}. \tag{3}$$

A potential of this functional form was applied in the past as a short-range component in phenomenological modelling of the quark-antiquark interaction [12]. The treatment of the potential, however, was so far restricted to the non-relativistic case described by the Schrödinger equation. Here, we show that the time-independent Dirac equation for this potential can be solved exactly. The solution is written in terms of linear combinations of the Hermite functions of a scaled and shifted argument. Based on this exact solution, we find a notable difference as compared to the Schrödinger case. We note that a potential of this functional form appears also in the graphene physics (e.g., the electrostatic potential caused by a gate voltage at the edge of a graphene strip [13]).

We consider the stationary one-dimensional Dirac equation for a spin $1/2$ particle of rest mass $m$ and energy $E$:

$$H|\psi\rangle = E|\psi\rangle, \tag{4}$$

with Hamiltonian $H = K + \Pi$, where $K$ is the kinetic energy operator, and the interaction operator $\Pi$ stands for the potential of the external field. With the general $2\times 2$ Hermitian potential matrix given by real functions $V, U, W, S$:

$$\Pi = V(x)\sigma_0 + U(x)\sigma_1 + W(x)\sigma_2 + S(x)\sigma_3 = \begin{pmatrix} V+S & U-iW \\ U+iW & V-S \end{pmatrix}, \tag{5}$$

where $\sigma_0$ is the unit matrix and $\sigma_{1,2,3}$ are the Pauli matrices, equation (4) for the two-component spinor wave function $\psi = (\psi_A, \psi_B)^T$ reads

$$\begin{pmatrix} V+S+mc^2 & -ic\hbar\frac{d}{dx}+U-iW \\ -ic\hbar\frac{d}{dx}+U+iW & V-S-mc^2 \end{pmatrix} \begin{pmatrix} \psi_A \\ \psi_B \end{pmatrix} = E \begin{pmatrix} \psi_A \\ \psi_B \end{pmatrix}, \tag{6}$$

where $c$ is the speed of light and $\hbar$ is the reduced Planck's constant (we use dimensional variables since this is useful for construction of new solutions using the known ones – see examples below). Without affecting the physical results, one can eliminate $U$ using the phase transformation $\psi \to e^{-i\varphi}\psi$ with $c\hbar \cdot d\varphi/dx = U$. Hence, we assume $U = 0$. The resulting system can be solved for several configurations of functions $V, W, S$ of the form $a + b/\sqrt{x}$.

## 2. Solution for a basic field configuration

A basic field configuration we consider is

$$(V,W,S) = \left(V_0 + \frac{V_1}{\sqrt{x}}, W_0, S_0 + \frac{S_1}{\sqrt{x}}\right) \quad (7)$$

with arbitrary $V_{0,1}, W_0, S_{0,1}$. Though this potential is defined on the positive semi-axis $x > 0$, one can extend it to the whole axis $x \in (-\infty, +\infty)$ by replacing $x$ by $|x|$. The solution for $x < 0$ is then readily constructed by replacing $x \to -x$, $c\hbar \to -c\hbar$ in the solution for $x > 0$.

To solve the Dirac equation for potential (7), we apply the Darboux transformation

$$\psi_A = a_1 \frac{dw}{dx} + a_2 w, \quad \psi_B = b_1 \frac{dw}{dx} + b_2 w \quad (8)$$

to reduce the system to a single second-order differential equation for a new variable $w(x)$. This is achieved by putting

$$(a_1, b_1) = \left(\sqrt{V_1 - S_1}, \sqrt{V_1 + S_1}\right), \quad (9)$$

$$a_2 = \frac{ib_1\left(E + mc^2 + S_0 - V_0 + (S_1 - V_1)F(x)\right) + a_1 W_0}{c\hbar}, \quad (10)$$

$$b_2 = -\frac{ia_1\left(-E + mc^2 + S_0 + V_0 + (S_1 + V_1)F(x)\right) + b_1 W_0}{c\hbar}. \quad (11)$$

where $F = 1/\sqrt{x}$. The resulting equation reads

$$\frac{d^2 w}{dx^2} + \frac{\left(A + B F(x) + (V_1^2 - S_1^2)F(x)^2 - ic\hbar\sqrt{V_1 - S_1}\sqrt{V_1 + S_1}F'(x)\right)}{c^2\hbar^2} w = 0, \quad (12)$$

where the prime denotes differentiation and

$$A = (E - V_0)^2 - (mc^2 + S_0)^2 - W_0^2, \quad (13)$$

$$B = -2\left((E - V_0)V_1 + (mc^2 + S_0)S_1\right). \quad (14)$$

In the spin and pseudo-spin symmetry cases, i.e. when $S_1^2 = V_1^2$, the terms proportional to $F(x)^2$ and $F'(x)$ vanish and equation (12) is reduced to a Schrödinger-like equation for the inverse-square-root potential:

$$\frac{d^2 w}{dx^2} + \frac{1}{c^2\hbar^2}\left(A + \frac{B}{\sqrt{x}}\right)w = 0 \quad (15)$$

with energy $A/(2mc^2)$ and potential $-B/(2mc^2\sqrt{x})$.

The general solution of this equation is presented in [14]. Rewritten in terms of our parameters, a fundamental solution for real $A$ and $B$ can be presented as

$$w = e^{-y^2/2}\left(H_\nu(y) - \text{sgn}(AB)\sqrt{2\nu}H_{\nu-1}(y)\right), \tag{16}$$

where
$$y = \sqrt{\frac{\sqrt{-4A}}{c\hbar}}\left(\sqrt{x} + \frac{B}{2A}\right), \tag{17}$$

and
$$\nu = \frac{B^2}{4c\hbar(-A)^{3/2}}. \tag{18}$$

We note that a second independent fundamental solution can be constructed by the change $c\hbar \to -c\hbar$ in equations (17)-(18).

Discussing the general case of field configuration (7) for arbitrary parameters, a main result we report is that a fundamental solution of equation (12) is given as

$$w = e^{-y^2/2}\left(H_\nu(y) + gH_{\nu-1}(y)\right), \tag{19}$$

where
$$y = \sqrt{-2\alpha_2}\left(\sqrt{2}x + \frac{\alpha_1}{2\alpha_2}\right), \tag{20}$$

$$(\nu, g) = \left(-\frac{\alpha_1^2}{4\alpha_2} - \frac{V_1^2 - S_1^2}{2\alpha_2 c^2\hbar^2}, \frac{\alpha_1}{\sqrt{-2\alpha_2}} - i\frac{\sqrt{V_1 - S_1}\sqrt{V_1 + S_1}}{\sqrt{-\alpha_2}c\hbar}\right), \tag{21}$$

and
$$(\alpha_1, \alpha_2) = \left(\frac{-B}{c\hbar\sqrt{-2A}}, \frac{\sqrt{-A}}{2c\hbar}\right). \tag{22}$$

We note that, mathematically, this solution applies not only for real parameters but for arbitrary *complex* parameters $V_{1,2}, W_0, S_{1,2}$ as well as for arbitrary complex variable $x$. This observation may be useful if one discusses non-Hermitian generalizations of Hamiltonian (5).

This solution is derived by the reduction of equation (12) to the biconfluent Heun equation [15,16]

$$\frac{d^2u}{dz^2} + \left(\frac{\gamma}{z} + \delta + \varepsilon z\right)\frac{du}{dz} + \frac{\alpha z - q}{z}u = 0, \tag{23}$$

the solution of which can be expanded in terms of (generally non-integer order) Hermite functions of a scaled and shifted argument [17]:

$$u = \sum_{n=0}^{\infty} c_n H_{n+\gamma-\alpha/\varepsilon}\left(\sqrt{-\varepsilon/2}(z + \delta/\varepsilon)\right) \tag{24}$$

(see the definition of the non-integer order Hermite function in [18]).

Following the approach of [19,20], we apply the transformation $\psi = \varphi(z)u(z)$ with $z = \sqrt{2x}$ and $\varphi = e^{\alpha_1 z + \alpha_2 z^2}$ to show that for the field configuration (7) the parameters of the biconfluent Heun equation obey the equations

$$\gamma = -1, \quad q^2 - \delta q + \alpha = 0 \tag{25}$$

(mathematically, this means that the regular singularity of equation (23) at $z = 0$ is *apparent*). With this, the series (24) terminates on the second term thus resulting in a closed-form solution involving just two Hermite functions:

$$u = e^{\alpha_1 z + \alpha_2 z^2}\left(H_{\gamma-\alpha/\varepsilon}\left(\sqrt{-\frac{\varepsilon}{2}}\left(z+\frac{\delta}{\varepsilon}\right)\right) + \frac{\sqrt{-\varepsilon}(\delta-q)}{\sqrt{2}\alpha}H_{1+\gamma-\alpha/\varepsilon}\left(\sqrt{-\frac{\varepsilon}{2}}\left(z+\frac{\delta}{\varepsilon}\right)\right)\right). \tag{26}$$

Rewritten in terms of parameters of equation (12), this yields the solution (19)-(22).

As regards the *general* solution of equation (12), it can be written as

$$w_G = e^{-y^2/2}\left(\Phi + \frac{g}{2\nu}\frac{d\Phi}{dy}\right) \tag{27}$$

with

$$\Phi = c_1 \cdot H_\nu(y) + c_2 \cdot {}_1F_1\left(-\frac{\nu}{2};\frac{1}{2};y^2\right), \tag{28}$$

where $c_{1,2}$ are arbitrary constants and ${}_1F_1$ is the Kummer confluent hypergeometric function. We conclude this section by noting that in general the involved Hermite and hypergeometric functions are not polynomials because the index $\nu = \gamma - \alpha/\varepsilon$ is not an integer.

### 3. Another field-configuration.

The solution of the Dirac equation for several other field-configurations is constructed in the same way – via reduction to the biconfluent Heun equation (23). For instance, for the spin symmetric configuration $S(x) - V(x) = C_s = \text{const}$ equations (6) reduce to

$$\frac{d^2\psi_A}{dx^2} + \frac{1}{c^2\hbar^2}\left((E+mc^2+C_s)(E-mc^2-C_s-2V) - c\hbar\frac{dW}{dx} - W^2\right)\psi_A = 0, \tag{29}$$

$$\psi_B = \frac{i}{E+mc^2+C_s}\left(W\psi_A - c\hbar\frac{d\psi_A}{dx}\right). \tag{30}$$

For the field-configuration

$$(V, W, S) = \left(\frac{V_1}{\sqrt{x}}, \frac{W_1}{\sqrt{x}}, C_s + \frac{V_1}{\sqrt{x}}\right), \tag{31}$$

reducing equation (29) to the biconfluent Heun equation (23), we verify that the parameters of the resulting equation fulfill equations (25). As a result, we arrive at a fundamental solution for $\psi_A$ written as an irreducible linear combination of two Hermite functions:

$$\psi_A = e^{-y^2/2}\left(H_\nu(y) + gH_{\nu-1}(y)\right) \tag{32}$$

with
$$y = \sqrt{-2\alpha_2}\left(\sqrt{2}x + \frac{\alpha_1}{2\alpha_2}\right) \tag{33}$$

and (compare with (21)-(22))

$$(\nu, g) = \left(-\frac{\alpha_1^2}{4\alpha_2} + \frac{W_1^2}{2\alpha_2 c^2\hbar^2}, \frac{\alpha_1}{\sqrt{-2\alpha_2}} - \frac{W_1}{c\hbar\sqrt{-\alpha_2}}\right), \tag{34}$$

$$(\alpha_1, \alpha_2) = \left(\frac{\sqrt{2}V_1(mc^2 + C_s + E)}{c\hbar\sqrt{(mc^2+C_s)^2 - E^2}}, \frac{\sqrt{(mc^2+C_s)^2 - E^2}}{2c\hbar}\right). \tag{35}$$

For completeness, in this case the general solution of the Dirac equation involving two independent fundamental solutions can be written as

$$\psi_A = e^{-y^2/2}\left(\Phi + \frac{g}{2\nu}\frac{d\Phi}{dy}\right) \tag{36}$$

with
$$\Phi = c_1 \cdot H_\nu(y) + c_2 \cdot {}_1F_1\left(-\frac{\nu}{2}; \frac{1}{2}; y^2\right), \tag{37}$$

where $c_{1,2}$ are arbitrary constants.

We note that for the pseudo-spin symmetry configuration $S + V = C_p = \text{const}$ with $V(x)$ and $W(x)$ given by equation (31), a fundamental solution for $\psi_B$ is constructed by the formal change $(\psi_A, C_s) \to (\psi_B, C_p)$ and $(V_1, W_1, E) \to (-V_1, -W_1, -E)$ in equations (32)-(35).

### 4. Bound states

To construct bound states, we (i) extend the potential to the whole $x$-axis $x \in (-\infty, +\infty)$ by assuming the potential being of argument $|x|$ instead of $x$, (ii) demand the wave function to vanish at $x \to \pm\infty$, and (iii) demand the wave function to be continuous in the origin $x = 0$. The solution for $x < 0$ is readily constructed by noting that the Dirac system (6) for a potential depending on $|x|$ is not changed if one replaces $x \to -x$ and $c\hbar \to -c\hbar$. It then turns out that the continuity condition in the origin results in the equation $\psi_A(0)\psi_B(0) = 0$ [21]. This is an important observation stating that for the bound states either the upper component $\psi_A$ or the lower one $\psi_B$ should vanish in the origin. As a result, one gets two subsets of eigenvalues – the ones for which $\psi_A(0) = 0$ and the ones for which $\psi_B(0) = 0$. We note that the vanishing of the wave function in the origin is a necessary condition for the non-relativistic limit [22].

As an example, consider the spin symmetric field configuration

$$V = \frac{V_1}{\sqrt{|x|}}, \quad W = 0, \quad S = \frac{V_1}{\sqrt{|x|}}. \tag{38}$$

This is a specific configuration that belongs to both families (7) and (31). Both approaches work yielding the same result. If this configuration is viewed as a particular case of the field configuration (7) with $V_0 = W_0 = S_0 = 0$ and $S_1 = V_1$, we have

$$A = E^2 - m^2c^4, \quad B = -2(E + mc^2)V_1, \tag{39}$$

and the general solution of the Dirac equation (6) for $x > 0$ is written as

$$\psi_A = c_1 w + c_2 \tilde{w}, \tag{40}$$

$$\psi_B = \frac{-ic\hbar}{E + mc^2} \frac{d\psi_A}{dx}, \tag{41}$$

where $c_{1,2}$ are arbitrary constants, $w$ is given via parameters $A$ and $B$ by equations (16)-(18), and $\tilde{w}$ is constructed from $w$ by the change $c\hbar \to -c\hbar$. As already mentioned above, the solution $(\psi_A^-, \psi_B^-)$ for $x < 0$ is constructed by the further change $x \to -x$:

$$\psi_A^- = (c_3 w + c_4 \tilde{w})\big|_{x \to -x}, \tag{42}$$

$$\psi_B^- = \left(\frac{-ic\hbar}{E + mc^2} \frac{d\psi_A}{dx}\right)\bigg|_{x \to -x}. \tag{43}$$

The condition of vanishing the wave function at the infinity leads to the simplification $c_2 = c_4 = 0$ (we note that then $\psi_A, \psi_A^-$ are real and $\psi_B, \psi_B^-$ are imaginary if $c_1, c_3$ are chosen real). With this, the continuity of the wave function at the origin is achieved if

$$c_3 w\big|_{x \to -0} = c_1 w\big|_{x \to +0}, \tag{44}$$

$$c_3 \frac{dw}{dx}\bigg|_{x \to -0} = c_1 \frac{dw}{dx}\bigg|_{x \to +0}. \tag{45}$$

Vanishing of the determinant of this system presents the exact equation for energy spectrum. Since $w(-x) = w(x)$ and $w'(-x) = -w'(x)$, this equation is reduced to $w(0)w'(0) = 0$ or $\psi_A(0)\psi_B(0) = 0$. Thus, for bound states, either $\psi_A$ or $\psi_B$ should vanish in the origin.

Consider the case $\psi_A(0)$. It is readily shown that this condition is rewritten as

$$H_\nu\left(-\sqrt{2\nu}\right) + \sqrt{2\nu}\, H_{\nu-1}\left(-\sqrt{2\nu}\right) = 0. \tag{46}$$

This is the exact equation for a subset of energy eigenvalues. We note that this type of spectrum equations that involve two Hermite functions are faced in several other physical situations (see, e.g., [14,23]). For $\text{sgn}(AB) = -1$ this is exactly the equation encountered when solving the Schrödinger equation for the inverse-square-root potential [14]. It has been shown that the equation possesses a countable infinite set of discrete positive roots $\nu_n$, $n \in \mathbb{N}$. This set determines the bound-state energy eigenvalues. We note that all $\nu_n$ are not integers so that the bound-state wave-functions are not polynomials.

The calculation lines are as follows. Substituting equations (39), into equation (18), one arrives at the following cubic equation for energy $E_n$:

$$c^2\hbar^2 \left(E_n - mc^2\right)^3 \nu_n^2 + \left(E_n + mc^2\right)V_1^4 = 0. \tag{47}$$

The discriminant $D = -4c^2\hbar^2\nu_n^2 V_1^8 \left(27 m^2 c^6 \hbar^2 \nu_n^2 + V_1^4\right)$ of this equation is negative, hence, the cubic has only one real root [16]. This root is conveniently written through the parameter

$$\theta = \frac{V_1^2}{3^{3/2} m\hbar c^3} \frac{1}{\nu_n}. \tag{48}$$

The result reads
$$E_n = mc^2 + 3mc^2 |\theta|^{2/3} \frac{\left(\sqrt{\theta^2 + 1} - 1\right)^{2/3} - |\theta|^{2/3}}{\left(\sqrt{\theta^2 + 1} - 1\right)^{1/3}}. \tag{49}$$

Given that $\nu_n$ are known via equation (46), this is an exact expression.

To approximately solve equation (46), we note that the arguments and indexes of the involved Hermite functions $H_\nu(z)$ belong to the *left* transient layer for which $z \approx -\sqrt{2\nu + 1}$ [24]. Following the approach of [14], we divide equation (46) by $\sqrt{2\nu}\, H_{\nu-1}\left(-\sqrt{2\nu}\right)$ and apply the proper approximations [24] to show that the equation has solutions only if $\text{sgn}(AB) = -1$. The resulting approximation for the latter case reads

$$F \equiv 1 + \frac{H_\nu\left(-\sqrt{2\nu}\right)}{\sqrt{2\nu}\,H_{\nu-1}\left(-\sqrt{2\nu}\right)} \approx f(\nu)\left(\sin\left(\pi(\nu+1/6)\right) + \frac{D_0}{\nu^{2/3}}\sin\left(\pi(\nu-1/6)\right)\right). \quad (50)$$

Here $f(\nu)$ is a non-oscillatory function which does not adopt zero and $D_0$ is the constant

$$D_0 = \frac{\Gamma(1/3)}{12\sqrt[3]{3}\,\Gamma(2/3)} \approx 0.11. \quad (51)$$

This is a highly accurate approximation (see Fig. 1).

Thus, the exact equation (46) is accurately approximated as

$$\frac{\sin\left(\pi(\nu+1/6)\right)}{\sin\left(\pi(\nu-1/6)\right)} + \frac{D_0}{\nu^{2/3}} = 0. \quad (52)$$

Treating the second term of this equation as a perturbation leads to a simple, yet, highly accurate approximation

$$\nu_n \approx n - \frac{1}{6} + \frac{\sqrt{3}D_0}{2\pi(n-1/6)^{2/3}} - \frac{\sqrt{3}D_0^2}{4\pi(n-1/6)^{4/3}}, \quad n = 1, 2, 3, \ldots. \quad (53)$$

The relative error is less than $10^{-4}$ for all orders $n \geq 2$ and the absolute error exceeds $10^{-4}$ only for the first root with $n=1$.

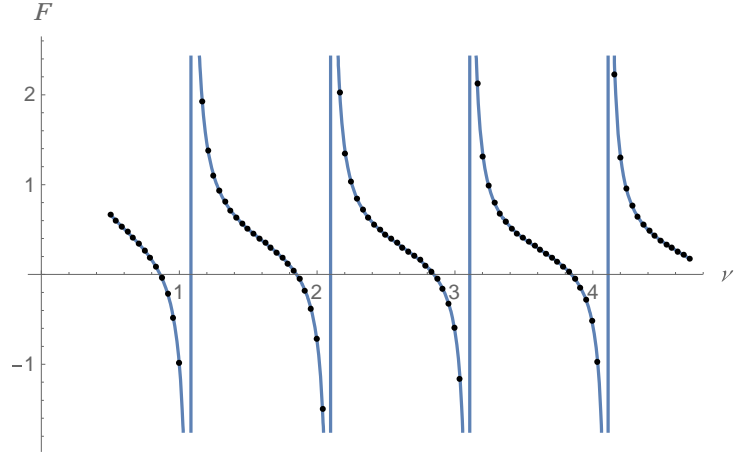

Fig. 1. Approximation (50) (filled circles) compared with the exact function $F$ (solid curves). For $\nu < 1/2$ the function does not possess roots.

With exact equation (49), keeping just the first term $\nu_n \approx n - 1/6$ in equation (53), the energy eigenvalues are expanded in terms of $n$ as

$$E_n \approx mc^2 \left(1 - \frac{2(3\lambda)^{2/3}}{(6n-1)^{2/3}} + \frac{2(3\lambda)^{4/3}}{(6n-1)^{4/3}}\right), \quad \lambda = \frac{V_1^2}{m\hbar c^3}. \quad (54)$$

This provides a good description of the whole sequence if $V_1^2/(m\hbar c^3) \leq 1$ (see Table 1).

| $n$ | 1 | 2 | 3 | 4 | 5 | 6 | 7 |
|---|---|---|---|---|---|---|---|
| $E_n$ (exact) | -0.07534 | 0.279904 | 0.438093 | 0.530429 | 0.592080 | 0.636685 | 0.670730 |
| $E_n$ (approx) | -0.08538 | 0.276806 | 0.436756 | 0.529710 | 0.591638 | 0.636390 | 0.670521 |

Table 1. Comparison of approximation (54) with the exact formula (49) for $V_1^2/(m\hbar c^3) = 1$: $(m, \hbar, c, V_1) = (1,1,1,-1)$.

Consider now the case $\psi_B(0)$. It is shown that this condition is rewritten as

$$H_\nu\left(-\sqrt{2\nu}\right) - \sqrt{2\nu} H_{\nu-1}\left(-\sqrt{2\nu}\right) = 0, \tag{55}$$

which differs from equation (46) only by the sign of the second term. Acting essentially in the same manner, we find that this equation is well approximated as

$$f(\nu)\left(\sin\left(\pi\nu - \frac{\pi}{6}\right) + \frac{1}{64\nu^{4/3}}\sin\left(\pi\nu + \frac{\pi}{6}\right)\right) = 0 \tag{56}$$

with $f(\nu) = \pi(2\nu)^{(3\nu+1)/6} e^{\nu/2}$. Neglecting the second term, we arrive at

$$\nu_n \approx n + \frac{1}{6}, \quad n = 0, 1, 2, 3, \ldots \tag{57}$$

(note that here $n$ runs starting from 0). The energy eigenvalues are expanded for large $n$ as

$$E_n \approx mc^2\left(1 - \frac{2(3\lambda)^{2/3}}{(6n+1)^{2/3}} + \frac{2(3\lambda)^{4/3}}{(6n+1)^{4/3}}\right). \tag{58}$$

Starting from $n=1$, this provides a rather good approximation if $V_1^2/(m\hbar c^3) \le 1$ (Table 2).

| $n$ | 0 | 1 | 2 | 3 | 4 | 5 | 6 |
|---|---|---|---|---|---|---|---|
| $E_n$ (exact) | -0.90450 | 0.073507 | 0.340973 | 0.472303 | 0.55273 | 0.607966 | 0.648679 |
| $E_n$ (approx) | -0.27567 | 0.078540 | 0.341908 | 0.472631 | 0.552883 | 0.608050 | 0.648731 |

Table 2. Comparison of approximation (58) with the exact formula (49) for $V_1^2/(m\hbar c^3) = 1$: $(m, \hbar, c, V_1) = (1,1,1,-1)$.

**5. Bound states for the case of electrostatic potential**

Consider the electrostatic potential

$$V = \frac{V_1}{\sqrt{|x|}}, \quad W = S = 0. \tag{59}$$

For $x > 0$, this is another particular case of the field configuration (7). Here

$$A = E^2 - m^2c^4, \quad B = -2EV_1 \tag{60}$$

and the general solution of the Dirac equation (6) is written as

$$\psi_A = \frac{dw_G}{dx} + \frac{i}{c\hbar}\left(E + mc^2 - \frac{V_1}{\sqrt{x}}\right)w_G, \qquad (61)$$

$$\psi_B = \frac{dw_G}{dx} + \frac{i}{c\hbar}\left(E - mc^2 - \frac{V_1}{\sqrt{x}}\right)w_G, \qquad (62)$$

where $w_G$ is the general solution (27),(28) and $y, \nu, g$ are given by equations (20)-(22). After some simplification, we have the result

$$\psi_A = e^{-\frac{y^2}{2}}\left((E + mc^2 - 2ic\hbar\alpha_2)\Phi + \frac{g}{2\nu}(E + mc^2 + 2ic\hbar\alpha_2)\frac{d\Phi}{dy}\right), \qquad (63)$$

$$\psi_B = e^{-\frac{y^2}{2}}\left((E - mc^2 - 2ic\hbar\alpha_2)\Phi + \frac{g}{2\nu}(E - mc^2 + 2ic\hbar\alpha_2)\frac{d\Phi}{dy}\right), \qquad (64)$$

where

$$\Phi = c_1 \cdot H_\nu(y) + c_2 \cdot {}_1F_1\left(-\frac{\nu}{2}; \frac{1}{2}; y^2\right). \qquad (65)$$

The requirement of vanishing of the wave function at $x \to +\infty$ gives the following linear relation between $c_1$ and $c_2$:

$$2^\nu \Gamma\left(\frac{\nu+1}{2}\right)c_1 + (-i)^\nu \sqrt{\pi} c_2 = 0. \qquad (66)$$

Proceeding in the same manner as in the previous case, we construct the solution of the Dirac equation (6) for $x < 0$ by replacing $x \to -x$ and $c\hbar \to -c\hbar$ in equations (20)-(22) and (63)-(65). Note that $c_1$ and $c_2$ should be replaced by new arbitrary constants, say, $c_3$ and $c_4$, respectively. It is then shown that the requirement of vanishing of the wave function at $x \to -\infty$ is satisfied if $c_4 = 0$. Finally, we verify that the requirement of continuity of the wave function at $x = 0$ results in the equation $\psi_A(0)\psi_B(0) = 0$. Hence, for bound states, either $\psi_A$ or $\psi_B$ should vanish in the origin.

Consider the case $\psi_A(0) = 0$. Since $c_4 = 0$, it is easier to use the solution for $x < 0$. Then, after some algebra, the equation $\psi_A(0) = 0$ is rewritten as (compare with (46) and (55))

$$H_\nu\left(-\frac{E}{mc^2}\sqrt{2\nu}\right) + \sqrt{2\nu}H_{\nu-1}\left(-\frac{E}{mc^2}\sqrt{2\nu}\right) = 0, \qquad (67)$$

where

$$\nu = \frac{m^2 c^3 V_1^2 / \hbar}{(m^2 c^4 - E^2)^{3/2}}. \qquad (68)$$

This is the exact equation for a subset of the energy spectrum. To treat this equation, we note that, since $|E/mc^2| < 1$, the indexes $\nu, \nu - 1$ and the argument $z = -\sqrt{2\nu} E/mc^2$ of the involved Hermite functions belong to the so called "inner region" for which $|z| < \sqrt{2\nu}$. One can then use the standard approximation for the Hermite function for this region [24]:

$$H_\nu(z) \propto 2^{\frac{1+\nu}{2}} e^{\frac{z^2-\nu+\nu \ln \nu}{2}} \left(1-\frac{z^2}{2\nu}\right)^{-1/4} \cos\left(\frac{\pi \nu}{2} - z\sqrt{\frac{\nu}{2}-\frac{z^2}{4}} - \frac{2\nu+1}{2}\arcsin\left(\frac{z}{\sqrt{2\nu}}\right)\right), \quad (69)$$

to arrive at the following highly accurate approximation:

$$\sin(\pi f) = 0, \quad (70)$$

where
$$f = \nu + \frac{1}{4} + \frac{\nu}{\pi}\left(\frac{E}{mc^2}\sqrt{1-\frac{E^2}{m^2c^4}} - \arccos\frac{E}{mc^2}\right). \quad (71)$$

The eigen-energies are thus defined as roots of the equation $f = k$, $k \in \mathbb{Z}$.

To get a general insight on the structure of the spectrum, it is useful to examine the behavior of $f$ as a function of energy, the latter being allowed to vary within the interval $E \in (-mc^2, mc^2)$ (Fig.2). Some characteristics of $f$ are

$$f|_{E \to -mc^2} = \frac{1}{4} + \frac{2\lambda}{3\pi} \equiv f_{\min}, \quad (72)$$

$$f|_{E=0} = \frac{1}{4} + \frac{\lambda}{2} \equiv f_0, \quad (73)$$

$$f|_{E \to +mc^2} \sim \nu + f_\infty, \quad f_\infty \equiv \frac{1}{4} - \frac{2\lambda}{3\pi}, \quad (74)$$

where
$$\lambda = \frac{V_1^2}{m\hbar c^3}. \quad (75)$$

The function starts from the minimal value $f_{\min}$ at $E = -mc^2$, adopts $f_0$ at $E = 0$, and diverges to plus infinity at $E \to +mc^2$. Since the function has a restricted variation range on the negative interval $E \in (-mc^2, 0)$, it is understood that there exists only a finite number of negative eigen-energies while the number of positive eigen-energies is infinite. The number of negative eigen-energies is exactly given as

$$n_- = \lfloor f_0 \rfloor - \lfloor f_{\min} \rfloor, \quad (76)$$

where $\lfloor .. \rfloor$ denotes the floor of a number.

For negative energies $E < 0$ the function $f(E)$ is well approximated by the parabola

$$f \approx f_0 + \frac{\lambda}{2}\left(\frac{E}{c^2 m}\right) + \frac{2\lambda}{3\pi}\left(\frac{E}{c^2 m}\right)^2, \tag{77}$$

while an appropriate approximation for positive energies $E > 0$ is

$$f \approx v + f_\infty + \frac{a\lambda^2}{v + b\lambda} \tag{78}$$

with

$$a = \frac{4 - 3\pi}{24\pi}, \quad b = \frac{3}{4}. \tag{79}$$

Note that $a$ is a small number: $a \approx -0.07$.

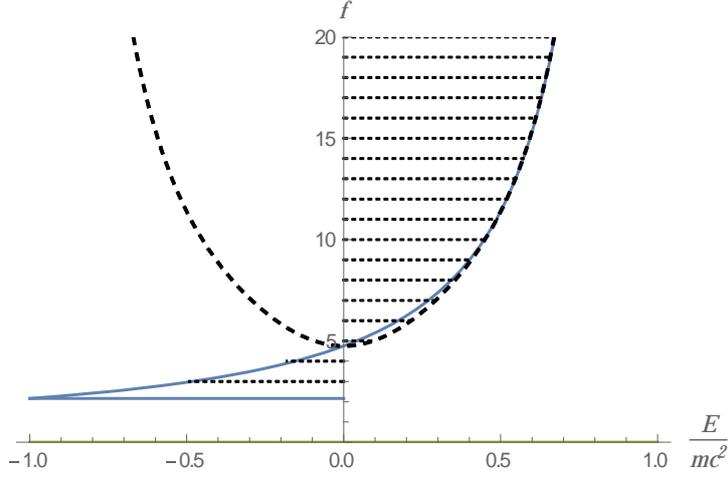

Fig. 2 Function $f(E)$ (solid line) for $(m, \hbar, c, V_1) = (1,1,1,-3)$. The intersections of horizontal dotted lines with $f(E)$ give the positions of energy levels $E_n$. There are only two negative eigen-energies for given parameters. The dashed line shows approximation (78).

With these approximations, one arrives at the following approximate spectrum. If the energy levels are numbered by a positive integer $n$ running from 1 to infinity, for negative energy levels $E_n < 0$ we have

$$E_n \approx -mc^2 \frac{3\pi}{8}\left(1 - \sqrt{1 + \frac{32}{3\pi}\frac{k - f_0}{\lambda}}\right), \quad k = n + \lfloor f_{\min} \rfloor, \tag{80}$$

where $n$ runs from 1 to $n_-$. For positive energy levels $E_n > 0$, using equation (68), we have

$$E_n \approx +mc^2 \sqrt{1 - \lambda^{2/3}/v^{2/3}}, \tag{81}$$

$$v = \frac{1}{2}\left(k - f_\infty - b\lambda + \sqrt{(k - f_\infty + b\lambda)^2 - 4a\lambda^2}\right), \quad k = n + \lfloor f_{\min} \rfloor, \tag{82}$$

where $n$ runs from $n_- + 1$ to infinity. This is a rather accurate result for all orders $n$ and for any $V_1$. Starting from a few lowest energy levels, the relative error is of the order or less than $10^{-3}$. The comparison of approximation (80) with the exact result for $V_1^2/(m\hbar c^3)=1$ is shown in Table 3.

| $n$ | 1 | 2 | 3 | 4 | 5 | 6 | 7 |
|---|---|---|---|---|---|---|---|
| $E_n$ | 0.297679 | 0.618900 | 0.723684 | 0.777986 | 0.811903 | 0.835392 | 0.852768 |
| $E_{approx}$ | 0.293394 | 0.611538 | 0.720164 | 0.775963 | 0.810589 | 0.834467 | 0.852079 |

Table 3. Comparison of approximation (81) with exact result for $(m,\hbar,c,V_1)=(1,1,1,-1)$.

We note that for large $n \to \infty$ the spectrum behaves as

$$E_n \approx +mc^2\sqrt{1-\frac{\lambda^{2/3}}{\left(n+\lfloor f_{\min}\rfloor - f_\infty\right)^{2/3}}}, \qquad (83)$$

This result, which is asymptotically exact, indicates that the Maslov index $\mu = -\{f_\infty\}$ (fractional part of the correction to $n$ [25,26]) depends on the potential's strength $V_1$. This is a notable feature that differs the Dirac case from the Schrödinger one. We recall that in the Schrödinger case the Maslov index is $\mu = -1/6$ [14].

The bound states corresponding to the case when the upper component of the wave function is even and the lower component is odd, that is when $\psi_B(0)=0$, are treated in the same manner. The exact equation for the second subset of the energy spectrum corresponding to this case is reduced to

$$H_\nu\left(-\frac{E}{mc^2}\sqrt{2\nu}\right) - \sqrt{2\nu}\,H_{\nu-1}\left(-\frac{E}{mc^2}\sqrt{2\nu}\right) = 0, \qquad (84)$$

which differs from equation (67) only by the sign of the second term. An accurate approximation of this equation reads

$$\sin(\pi f) = 0 \qquad (85)$$

with
$$f = \nu - \frac{1}{4} + \frac{\nu}{\pi}\left(\frac{E}{mc^2}\sqrt{1-\frac{E^2}{m^2c^4}} - \arccos\frac{E}{mc^2}\right), \qquad (86)$$

which differs from equation (71) only by the sign of 1/4. Doing the same steps as in the previous case, we arrive at the approximate spectrum given by the same equations (80)-(82) with $f_{\min}, f_0, f_\infty$ modified such as $1/4$ is replaced by $-1/4$. The two subsets of energy levels for $V_1^2/(m\hbar c^3)=1$ corresponding to the cases $\psi_A(0)=0$ and $\psi_B(0)=0$ are compared in Table 4. As seen, the second subset possesses an additional negative level $E \approx -0.965886$, the other levels being located between two adjacent levels of the first subset.

| $n$ | 1 | 2 | 3 | 4 | 5 | 6 | 7 |
|---|---|---|---|---|---|---|---|
| $E_n$, eq.(67) | 0.297679 | 0.618900 | 0.723684 | 0.777986 | 0.811903 | 0.835392 | 0.852768 |
| $E_n$, eq.(84) | –0.96589 | 0.495364 | 0.674916 | 0.751128 | 0.794616 | 0.823205 | 0.843645 |

Table 4. Exact energy levels corresponding to the cases $\psi_A(0) = 0$ and $\psi_B(0) = 0$, $\lambda = 1$.

**Acknowledgments**

This research was supported by the Russian-Armenian University at the expense of the Ministry of Education and Science of the Russian Federation, the Armenian Science Committee (SC Grant 18T-1C276), and the Armenian National Science and Education Fund (ANSEF Grant No. PS-5701).

**References**

1. W. Greiner, *Relativistic Quantum Mechanics. Wave equations* (Springer, Berlin, 2000).
2. V.G. Bagrov and D.M. Gitman, *The Dirac Equation and its Solutions*, (de Gruyter, Boston 2014)
3. P.A. Cook, "Relativistic harmonic oscillators with intrinsic spin structure", Lett. Nuovo Cimento **1**, 419-426 (1971)
4. M. Moshinsky and A. Szczepaniak, "The Dirac oscillator", J. Phys. A **22**, L817-L819 (1989).
5. P. Kennedy, "The Woods-Saxon potential in the Dirac equation", J. Phys. A **35**, 689-698 (2002).
6. J.Y. Guo, Y. Yu, and S.W. Jin, "Transmission resonance for a Dirac particle in a one-dimensional Hulthén potential", Cent. Eur. J. Phys. **7**, 168-174 (2009).
7. A. Kratzer, "Die ultraroten Rotationsspektren der Halogenwasserstoffe", Z. Phys. **3**, 289-307 (1920).
8. E. Schrödinger, "Quantisierung als Eigenwertproblem (Erste Mitteilung)", Annalen der Physik **76**, 361-376 (1926).
9. P.M. Morse, "Diatomic molecules according to the wave mechanics. II. Vibrational levels", Phys. Rev. **34**, 57-64 (1929).
10. G. Pöschl, E. Teller, "Bemerkungen zur Quantenmechanik des anharmonischen Oszillators", Z. Phys. **83**, 143-151 (1933).
11. C. Eckart, "The penetration of a potential barrier by electrons", Phys. Rev. **35**, 1303-1309 (1930).
12. X. Song and H. Lin, "A new phenomenological potential for heavy quarkonium", Z. Phys. C Particles and Fields **34**, 223-231 (1987).
13. P.G. Silvestrov and K.B. Efetov, "Charge accumulation at the boundaries of a graphene strip induced by a gate voltage: Electrostatic approach", Phys. Rev. B **77**, 155436 (2008).
14. A.M. Ishkhanyan, "Exact solution of the Schrödinger equation for the inverse square root potential $V_0/\sqrt{x}$", Eur. Phys. Lett. **112**, 10006 (2015).

15. A. Ronveaux (ed.), *Heun's Differential Equations* (Oxford University Press, London, 1995).
16. F.W.J. Olver, D.W. Lozier, R.F. Boisvert, and C.W. Clark (eds.), *NIST Handbook of Mathematical Functions* (Cambridge University Press, New York, 2010).
17. T.A. Ishkhanyan and A.M. Ishkhanyan, "Solutions of the bi-confluent Heun equation in terms of the Hermite functions", Ann. Phys. **383**, 79-91 (2017).
18. N.N. Lebedev and R.R. Silverman, *Special Functions and Their Applications* (Dover Publications, New York, 1972).
19. A. Ishkhanyan and V. Krainov, "Discretization of Natanzon potentials", Eur. Phys. J. Plus **131**, 342 (2016).
20. A.M. Ishkhanyan, "Schrödinger potentials solvable in terms of the general Heun functions", Ann. Phys. **388**, 456-471 (2018).
21. R.L. Hall and P. Zorin, "Nodal theorems for the Dirac equation in $d \geq 1$ dimensions", Ann. Phys. (Berlin) **526**, 79-86 (2014).
22. M. Znojil, "Comment on "Conditionally exactly soluble class of quantum potentials", Phys. Rev. A **61**, 066101 (2000).
23. A.S. de Castro, "Comment on "Fun and frustration with quarkonium in a 1+1 dimension," by R.S. Bhalerao and B. Ram, Am. J. Phys. **69**, 817-818 (2001)", Am. J. Phys. **70**, 450-451 (2002).
24. G. Szegö, *Orthogonal Polynomials*, 4th ed. (Amer. Math. Soc., Providence, 1975).
25. A.M. Ishkhanyan and V.P. Krainov, "Maslov index for power-law potentials", JETP Lett. **105**, 43-46 (2017).
26. C. Quigg and J.L. Rosner, "Quantum mechanics with applications to quarkonium", Phys. Rep. **56**, 167-235 (1979).